\documentclass[11pt,a4paper]{article}

\usepackage[T1]{fontenc}
\usepackage[osf,sc]{mathpazo}   
\usepackage{geometry}
\usepackage{amsmath}
\usepackage{amssymb}
\usepackage{booktabs}
\usepackage{textcomp}
\newcommand{\ct}{\textcent}  
\usepackage{graphicx}
\usepackage{hyperref}
\hypersetup{colorlinks=true, linkcolor=black, citecolor=black, urlcolor=blue}

\newcommand{\Os}{\Omega^{\!*}}
\newcommand{\G}{\mathrm{G}}

\title{An Asymmetric Formula for Interval Consonance \\ and its Relation to Harmonic Coincidence}
\author{David De Roure%
  \thanks{Department of Engineering Science, University of Oxford, Oxford, UK.
  Also Technical Director, Centre for Practice \& Research in Science \&
  Music (PRiSM), Royal Northern College of Music, Manchester, UK.
  \texttt{david.deroure@oerc.ox.ac.uk}}}
\date{June 2026}

\begin{document}
\maketitle


\begin{abstract}
Euler's \emph{Gradus Suavitatis} (1739) assigns a dissonance value to a musical
interval $p/q$ by the formula $\G(p/q) = 1 + \Os(p) + \Os(q)$, where
$\Os(n) = \sum_i e_i(p_i - 1)$ sums the weighted prime exponents of~$n$.
We propose the simpler asymmetric formula $f(p/q) = p + \Os(q)$, which
treats numerator and denominator differently and performs comparably on
standard consonance data.
We also show that, under a model in which harmonics are integer-indexed and
counted uniformly up to a fixed truncation level, Gradus is equivalent to a
weighted harmonic coincidence count with weights $w(n) = \Os(n)$,
connecting it to Galileo's earlier pulse-coincidence model (1638).
The formula naturally generates a coprime integer triangle $T(n,k) = n + \Os(k)$,
whose rightmost diagonal gives the two-stage dissonance of the superparticular
(consecutive-harmonic) intervals.
The formula $f$ admits a simple two-stage interpretation in terms of
harmonic context and partial recognition, which we offer as a speculative
perceptual hypothesis.
\end{abstract}


\section{Background}

Euler's Gradus Suavitatis \cite{euler1739}, catalogued by the author in OEIS A275314 \cite{oeis}, assigns to each musical
interval a positive integer measuring its arithmetic dissonance. Throughout
this paper, $\Os(n)$ denotes the completely additive arithmetic function
\[
  \Os(n) = \sum_i e_i(p_i - 1),
\]
where $n = p_1^{e_1} p_2^{e_2} \cdots$ is the prime factorisation of $n$,
with $\Os(1) = 0$. For a ratio $p/q$ in lowest terms, the characteristic
number is $n = p \cdot q$, and:
\[
  \G(p/q) = 1 + \Os(p \cdot q)
\]
Since $\Os$ is completely additive, $\G(p/q) = 1 + \Os(p) + \Os(q)$, treating
numerator and denominator symmetrically. The weights $(p_i - 1)$ for prime
$p_i$ encode Euler's aesthetic judgment that large primes represent greater
``difficulty.''

Euler's formula correctly ranks the 13 standard intervals of Western music
against human consonance ratings \cite{krumhansl1990} with Spearman
$\rho = 0.979$. We note it is exactly tied on the major third/major sixth pair
and on a three-way tie at Gradus~8, while these are distinct values in the
human consonance rank.

The 13 intervals are the standard dyads of the Western chromatic and diatonic
scales shown in Table~\ref{tab:intervals}. The human ratings follow
Krumhansl \cite{krumhansl1990}, which reports data from musically trained
Western listeners and assumes octave equivalence. With only 13 data points,
rank correlations should be read as illustrative comparisons rather than
statistically decisive results; the ordering is well-replicated
\cite{lahdelma2016,lahdelma2021} and broadly consistent with cross-cultural
data \cite{mcdermott2016,bowling2018}, though precise values depend on
musical training and cultural exposure.

\medskip\noindent
\textbf{Organisation.}
Section~2 shows that Gradus admits, within a discrete harmonic model, an
exact reinterpretation as a weighted harmonic coincidence count. Section~3 introduces $\max(p,q)$ and its four
classical interpretations. Section~4 presents the asymmetric formula
$f(p/q) = p + \Os(q)$. Section~5 compares all three formulae in detail.
Section~6 examines the one tie that neither $f$ nor $\max$ can resolve.
Section~7 proposes a speculative two-stage perceptual interpretation of $f$.
Section~8 surveys the consonance landscape, established scales, and equal
temperament systems. Section~9 develops a
partial-beating tolerance model that connects the arithmetic formula to
acoustic roughness and register dependence, and reveals a duality between
the perceptual cost of an interval and its sensitivity to mistuning.
Section~10 presents the triangle $T(n,k)$ and its companion OEIS sequences
A397104 and A397106, and concludes.


\section{Euler's Gradus as Harmonic Coincidence}

Throughout this paper the harmonic series is treated as a discrete,
integer-indexed model: harmonics are indexed by positive integers $m$,
counted uniformly, and spectral envelopes, inharmonicity, and equal
temperament deviations are set aside. The derivation below is mathematically
exact within this idealisation; its applicability to real musical sounds ---
which have finite and instrument-dependent spectra, slight inharmonic
stretch, and other deviations from the ideal --- is the empirical question
that the correlation data address.

For two notes at frequency ratio $p:q$ (coprime), the $m$-th harmonic
coincidence occurs at absolute frequency $mp \cdot f_0$, which is the
$(mp)$-th harmonic of the lower note and the $(mq)$-th harmonic of the upper
note. The following derivation assumes a fixed truncation level $M$, the same
for all intervals, and that each coincidence event contributes equally to the
score. In practice $M$ is determined by the spectral content of the sound and
the frequency range of hearing, and may vary with interval; the identity below
is therefore a model-dependent result, not a universal physical law.

Assigning weight $w(n) = \Os(n)$ to the $n$-th harmonic of each note,
the weighted contributions to the lower and upper note respectively, summed
over $M$ coincidences, are:
\begin{align*}
  \text{Lower:} &\quad \sum_{m=1}^{M} \Os(m p) = \Os(p)\cdot M + C(M) \\
  \text{Upper:} &\quad \sum_{m=1}^{M} \Os(m q) = \Os(q)\cdot M + C(M)
\end{align*}
where $C(M) = \sum_{m=1}^{M} \Os(m)$ is independent of both $p$ and $q$.
Both identities follow from the complete additivity of $\Os$:
$\Os(mn) = \Os(m) + \Os(n)$ for all positive integers $m, n$.
This is an arithmetic identity, not a physical one: it is because $\Os$ is
completely additive that the multiplicative structure of prime factorisation
and the additive structure of harmonic series indexing happen to be compatible.
This arithmetic coincidence is what makes the model tractable; it is a
modelling bridge rather than a physical necessity.
Adding the two sums gives the symmetric score:
\[
  \text{Score} = M \cdot \bigl(\Os(p) + \Os(q)\bigr) + 2\,C(M)
               = M \cdot (\G(p/q) - 1) + 2\,C(M)
\]

\textbf{Gradus therefore admits, within this model, an exact reinterpretation
as a weighted harmonic coincidence count}, with harmonics weighted by their
own prime complexity. This confirms the physical intuition behind Euler's formula
but reveals its self-referential structure: recovering Gradus from
coincidences requires weights that are themselves Gradus-like. We emphasise
that this is a reinterpretation of Gradus rather than Euler's own
motivation, which was aesthetic and number-theoretic.

Unweighted coincidence counting (flat weights) achieves $\rho = 0.791$ across
all coprime ratios with $p, q \leq 16$. The weight function $\log n$ achieves
$\rho = 0.795$ without any prime structure, establishing a ceiling for
physics-based models that do not invoke prime factorisation.


\section{Galileo's Model and \texorpdfstring{$\max(p,q)$}{max(p,q)}}

Galileo \cite{galilei1638, barbieri2001} described consonance in terms of pulse coincidences: two strings
at ratio $p:q$ produce coinciding pulses every $p$ vibrations of the lower
string. The ``sweetness'' of the interval grows with the frequency of
coincidences, i.e.\ inversely with~$p$. The metric $\max(p,q) = p$ is
therefore:

\begin{enumerate}
  \item \textbf{Galileo's pulse model} --- vibrations between coincidences;
  \item \textbf{Harmonic series position} --- the upper note is partial $p$ of
        the implied fundamental \cite{rameau1722};
  \item \textbf{Farey index} --- the smallest $n$ such that $q/p$ appears in
        the Farey sequence $F_n$ (Farey 1816; proved by Cauchy);
  \item \textbf{First coincidence harmonic} --- the position in the lower
        note's series where the two notes first meet.
\end{enumerate}

These four descriptions, discovered independently across four centuries, are
mathematically identical. $\max(p,q)$ achieves $\rho = 0.989$ against human
ratings, marginally exceeding Gradus, but cannot distinguish intervals with
the same numerator (e.g.\ $5/4$ and $5/3$ both have $\max = 5$).

Galileo's account predates Euler's \emph{Gradus Suavitatis} by a century, and
is physical rather than arithmetic: it concerns the periodic coincidence of
string vibrations, not the factorisation of integers. It is therefore
significant that the purely arithmetic measure $\max(p,q)$, and more so the
formula $f(p/q) = p + \Os(q)$ of Section~4, reproduce the same ordering as
Galileo's dynamical model. This suggests that prime-arithmetic complexity and
vibration period may be mathematically related in ways that are not yet fully
understood, and that the parallel between physics and arithmetic here is
unlikely to be coincidental.


\section{A Simpler Asymmetric Formula}

Throughout this paper, $p/q$ denotes a frequency ratio in lowest terms with
$p > q \geq 1$ and $\gcd(p,q) = 1$, so that $p/q > 1$ represents an
interval above unison and $p$ is always the numerator (upper note).

Euler's formula is symmetric: $\Os(p) + \Os(q)$. Human ratings, however,
distinguish numerator and denominator --- the bass note and the upper note
play different perceptual roles. We find empirically that the asymmetric
formula:
\[
  \boxed{f(p/q) = p + \Os(q)} \qquad (p > q \geq 1,\ \gcd(p,q)=1)
\]
achieves $\rho = 0.989$ on the 13-interval dataset, matching $\max(p,q)$ and
slightly exceeding Gradus. The formula differs from Gradus only in replacing
$\Os(p)$ with~$p$ for the numerator.

The formula naturally generates a coprime integer triangle
\[
  T(n,k) = n + \Os(k), \qquad 1 \leq k \leq n,\ \gcd(n,k) = 1,
\]
whose rightmost diagonal $T(n,\,n-1)$ gives the two-stage dissonance of the
superparticular (consecutive-harmonic) intervals $n/(n-1)$. Both the triangle
and its diagonal are treated in detail in Section~10.

The choice of $\Os(q)$ for the denominator term is motivated by its role in
Euler's Gradus and by the correspondence with human data; however, other
complexity measures of $q$ --- for example $\log q$, the unweighted prime
count $\Omega(q)$, or the largest prime factor --- yield structurally similar
but numerically distinct formulae.

\medskip
\noindent The formula $f(p/q) = p + \Os(q)$ decomposes as the sum of two
quantities:

\medskip
\begin{tabular}{lll}
\toprule
Component & Formula & Meaning \\
\midrule
Upper note reach & $p$ & Partial number of upper note in implied harmonic series \\
Bass context depth & $\Os(q)$ & Prime complexity of lower note's position in series \\
\bottomrule
\end{tabular}

\medskip\noindent
Section~7 proposes a perceptual hypothesis behind this decomposition.


\section{Comparison of the Three Formulae}

Table~\ref{tab:intervals} invites a direct comparison of the three arithmetic
formulae; several observations follow.

\begin{table}[h]
\centering
\caption{Dissonance values for the 13 standard just intervals, ordered by
human consonance rank \cite{krumhansl1990} (rank 1 = most consonant).}
\label{tab:intervals}
\smallskip
\begin{tabular}{llrrrrr}
\toprule
Interval & Ratio & $\G(p/q)$ & $f(p/q)$ & $\max(p,q)$ & $H(p/q)$ & Human rank \\
\midrule
Unison        & $1/1$   &  1 &  1 &  1 &     1 &  1 \\
Octave        & $2/1$   &  2 &  2 &  2 &     2 &  2 \\
Fifth         & $3/2$   &  4 &  4 &  3 &     6 &  3 \\
Fourth        & $4/3$   &  5 &  6 &  4 &    12 &  4 \\
Major third   & $5/4$   &  7 &  7 &  5 &    20 &  5 \\
Minor third   & $6/5$   &  8 & 10 &  6 &    30 &  6 \\
Major sixth   & $5/3$   &  7 &  7 &  5 &    15 &  7 \\
Minor sixth   & $8/5$   &  8 & 12 &  8 &    40 &  8 \\
Major second  & $9/8$   &  8 & 12 &  9 &    72 &  9 \\
Minor seventh & $9/5$   &  9 & 13 &  9 &    45 & 10 \\
Major seventh & $15/8$  & 10 & 18 & 15 &   120 & 11 \\
Minor second  & $16/15$ & 11 & 22 & 16 &   240 & 12 \\
Tritone       & $45/32$ & 14 & 50 & 45 &  1440 & 13 \\
\bottomrule
\end{tabular}
\smallskip

\noindent\small $H(p/q) = p \cdot q$ is the Tenney height \cite{tenney1988},
the standard complexity measure in just intonation theory.
Spearman $\rho$ against human rank: $G = 0.979$, $f = 0.989$,
$\max = 0.989$, $H = 0.978$.
\end{table}

\subsection*{Agreement on the extremes}
All three formulae agree without qualification on the most and least consonant
intervals. The unison (1), octave (2), fifth (3/2), and fourth (4/3) are
assigned the lowest values by every measure, consistent with their historical
priority in counterpoint and their prominence in non-Western musical systems.
The tritone (45/32) receives the highest value under every formula, and the
minor and major seconds cluster near the top of the dissonance scale in all
three cases. The disagreements are concentrated in the middle of the table,
among intervals that are also the most contested in historical music theory.

\subsection*{Correlation and ties}
Euler's Gradus achieves a Spearman rank correlation of $\rho = 0.979$ against
the human data; $f$ and $\max$ both achieve $\rho = 0.989$. The difference is
modest, and caution is warranted given the small sample (13 intervals) and the
cultural specificity of the rating data (see Section~6). More informative than
the correlation coefficient is the tie structure. Gradus produces two tie
groups among the 13 intervals: the major third and major sixth share $G = 7$,
and the minor third, minor sixth, and major second share $G = 8$. The
three-way tie at $G = 8$ is particularly awkward, as the three intervals span
three consecutive human ranks (6, 8, 9). The formula $f$ resolves this
three-way tie --- giving $f = 10$, $12$, $12$ respectively --- though it
introduces a two-way tie between the minor sixth and major second (both $f =
12$, human ranks 8 and 9). The metric $\max(p,q)$ shares the major third /
major sixth tie with $f$, but adds a further tie between the major second and
minor seventh (both $\max = 9$, human ranks 9 and 10). On the criterion of
tie-resolution among these 13 intervals, $f$ performs best, Gradus
intermediate, and $\max$ least well.

\subsection*{Dynamic range and the tritone}
The three formulae differ most sharply in their treatment of the tritone. Gradus
gives $G(45/32) = 14$, only four units above the major seventh ($G = 10$).
Both $f$ and $\max$ assign substantially higher values ($f = 50$, $\max = 45$),
placing the tritone in a category of its own and reflecting the large
numerator $p = 45$ that the asymmetric formulae weight directly. Whether this
wider dynamic range better represents human experience is difficult to assess
from rank-correlation data alone; what can be said is that the tritone is
consistently rated as by far the most dissonant interval in the human data,
and both $f$ and $\max$ capture that separation more emphatically than Gradus.

A caveat is necessary: the tritone used throughout this paper is $45/32$, the
5-limit augmented fourth generated within the diatonic scale as
$45/32 = (9/8) \cdot (5/4)$. This choice is conventional in 5-limit just
intonation but is not the only defensible one. The septimal tritone $7/5$
(582.5 cents) gives $f(7/5) = 7 + \Os(5) = 11$ and $G(7/5) = 11$ --- barely
more dissonant than the major third under either measure. The undecimal tritone
$11/8$ gives $f = 14$. These alternatives sit in different prime limits and
carry different musical connotations: $7/5$ is the tritone of barbershop and
some jazz practice; $45/32$ is the tritone of classical diatonic harmony.
The extreme values $f = 50$ and $\max = 45$ are therefore properties of the
5-limit diatonic tritone specifically, not of the tritone concept in general,
and should be interpreted accordingly. The high values are, however,
consistent with the 5-limit diatonic context in which the human rating data
were collected.

\subsection*{Symmetry}
Gradus is symmetric in $p$ and $q$: $G(p/q) = G(q/p)$. This reflects Euler's
focus on the product $p \cdot q$ as the relevant arithmetic quantity, and is
appropriate if the perceptual roles of the two notes are interchangeable.
Both $f$ and $\max$ are asymmetric, treating the numerator (upper note) and
denominator (lower note) differently, as is the case when one note functions
as a bass and the other as a melody note. The symmetry of Gradus means it
cannot distinguish $9/5$ from $16/9$ --- two minor sevenths that differ by
only 1 cent but arise from different tuning systems --- whereas $f$ gives
$f(9/5) = 13$ and $f(16/9) = 20$. Whether this asymmetry reflects a genuine
perceptual distinction or an artefact of the formula is an open question;
the perceptual hypothesis of Section~7 offers one possible motivation.

\subsection*{Interval inversion}
Two intervals whose frequency ratios multiply to 2 are \emph{inversions} of
each other within the octave: the major third (5/4) inverts to the minor
sixth (8/5), the fifth (3/2) to the fourth (4/3), and so on. Under Gradus
and under Tenney height, inversionally related intervals receive similar but
not identical values ($G(5/4) = 7$ vs $G(8/5) = 8$; $H(5/4) = 20$ vs
$H(8/5) = 40$). The formula $f$ is more strongly asymmetric: $f(5/4) = 7$
but $f(8/5) = 12$, a gap of 5 units. This captures the perceptual asymmetry
--- a minor sixth in root position (bass on the lower note) is generally
heard as less stable than a major third --- but amplifies it considerably.
Whether the magnitude of the asymmetry is empirically justified would require
testing beyond the Krumhansl dataset, which does not distinguish root-position
from inverted voicings.

\subsection*{Tenney height}
The Tenney height $H(p/q) = p \cdot q$ \cite{tenney1988}, also known as the
Benedetti height, is included in Table~\ref{tab:intervals} for comparison. It
achieves $\rho = 0.978$ against the human data, comparable to Gradus and
slightly below $f$ and $\max$. Tenney height is symmetric like Gradus, but
multiplicative rather than additive: it measures the prime complexity of the
product $p \cdot q$ without distinguishing numerator from denominator. Its
ordering of the 13 intervals differs from human judgments most sharply on the
major third / major sixth pair: $H(5/4) = 20 > H(5/3) = 15$, placing the
major sixth above the major third in consonance --- directly opposite to human
ratings. This failure is qualitatively different from the tie produced by $f$
and $\max$: all additive prime-weighted formulae tie $5/4$ and $5/3$ because
$\Os(4) = \Os(3)$, whereas Tenney actively inverts them. The result supports
the view that a multiplicative symmetric measure is not the right structure
for this problem.

\subsection*{Simplicity and interpretability}
The metric $\max(p,q)$ is strikingly simple, requires no prime factorisation,
and admits four independent mathematical characterisations (Section~3). These
properties make it an attractive baseline. Its weakness is precisely that
simplicity: it is insensitive to whether the denominator is a power of 2 or
a large prime, treating $4/3$ and $7/3$ identically in the denominator.
Gradus captures this prime sensitivity through $\Os(q)$ but applies the same
weighting to the numerator, where the evidence suggests a different treatment
is warranted. The formula $f$ retains prime sensitivity in the denominator
while replacing the numerator term with the raw partial number, as is
motivated by the perceptual hypothesis of Section~7. Each formula thus represents a
different theory of what makes an interval consonant, and the empirical data
do not decisively separate them.

\subsection*{Sensitivity near just ratios}
\begin{figure}[h]
\centering
\includegraphics[width=\textwidth]{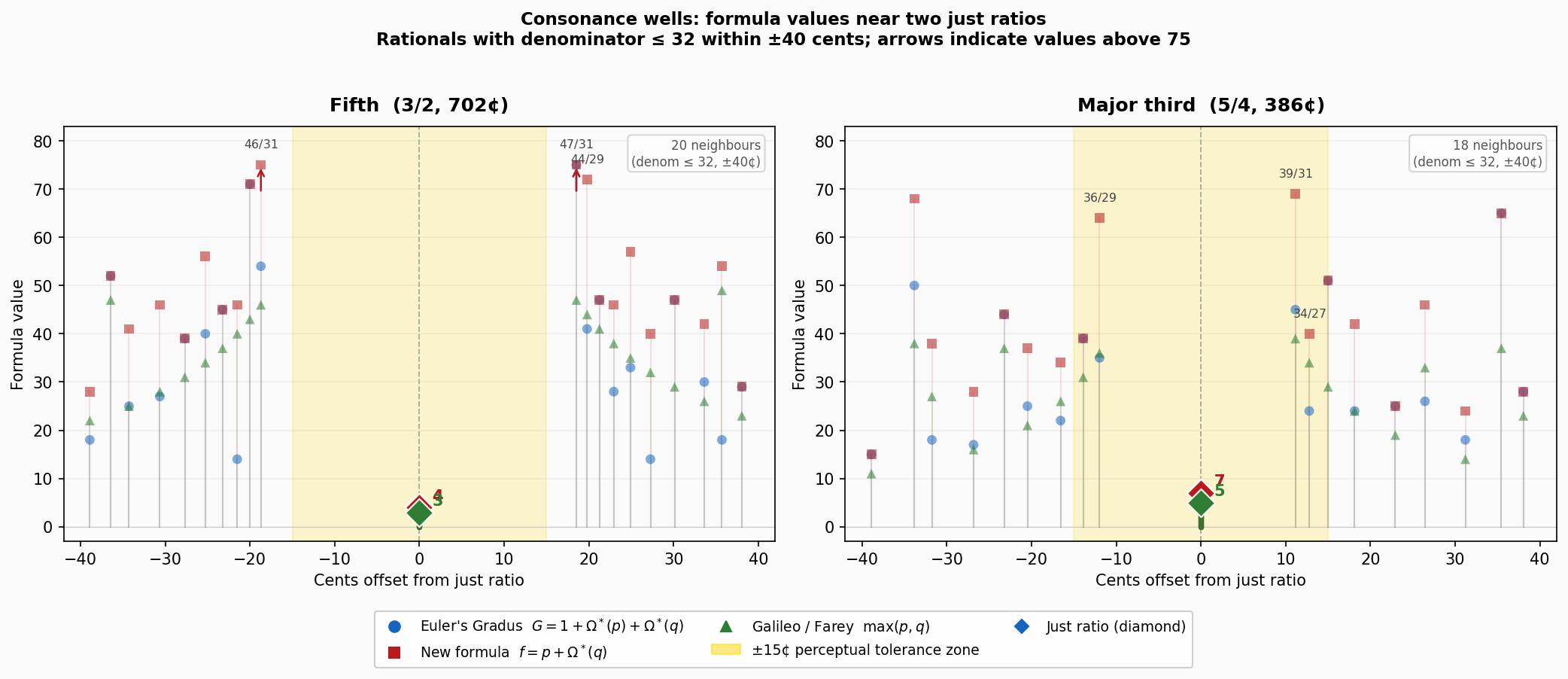}
\caption{Formula values for all rationals with denominator $\leq 32$ within
$\pm 40$ cents of the just fifth (3/2, left) and major third (5/4, right).
Diamond markers show the just ratio; circles, squares, and triangles show
neighbouring rationals under $G$, $f$, and $\max$ respectively. Upward
arrows indicate values above 75. The gold band marks the $\pm 15$ cent
perceptual tolerance zone; no rival rational falls within it for either
interval.}
\label{fig:wells}
\end{figure}

Figure~\ref{fig:wells} illustrates these differences directly. For two
showcase intervals --- the fifth (3/2) and the major third (5/4) --- every
rational number with denominator $\leq 32$ within $\pm 40$ cents of the just
ratio is plotted against its formula value. The just ratio sits at the centre
of a deep trough, surrounded by rationals with substantially higher values
under all three formulae. The red markers ($f$) are consistently the tallest:
moving roughly 20 cents from the just fifth, $f$ reaches values of 70 or
above while Gradus ($G$, blue) remains in the 40--55 range. The green markers
($\max$) are intermediate but track the numerator $p$ directly and so cluster
at values that increase smoothly with denominator size. Within the $\pm 15$
cent tolerance zone (gold shading), no rival rational with denominator $\leq
32$ exists for either interval, confirming that the just ratios are genuinely
isolated at low complexity in their neighbourhood.

\subsection*{Choice of formula}
The three formulae are more complementary than competing. For a context
requiring only a rough ordering of standard intervals, any of the three
suffices. Where fine distinctions matter --- competing tuning systems,
intervals outside the standard 13, or analysis of the consonance landscape
under equal temperament --- the formulae diverge in informative ways, and the
choice should be guided by the theoretical question at hand. Harrison \&
Pearce \cite{harrisonpearce2020} provide a comprehensive computational
framework decomposing simultaneous consonance into roughness, harmonicity,
and cultural familiarity components; the formula $f(p/q)$ may be understood
as an arithmetic approximation to the harmonicity component of that
decomposition, with the partial-beating tolerance model of Section~9
corresponding to the roughness component.

\medskip
\begin{tabular}{lll}
\toprule
Formula & $\rho$ vs.\ human & Notes \\
\midrule
Tenney height: $H = p \cdot q$          & 0.978 & Symmetric; inverts 5/4 vs 5/3 \\
Euler's Gradus: $1 + \Os(p) + \Os(q)$ & 0.979 & Symmetric; two tie groups \\
Galileo / $\max(p,q)$                  & 0.989 & Simplest; one tie \\
$f(p/q) = p + \Os(q)$                  & \textbf{0.989} & Asymmetric; two-stage hypothesis \\
Roughness (Plomp--Levelt)              & 0.709--0.863 & Register-dependent; fails on tritone$^*$ \\
\bottomrule
\end{tabular}

\medskip
\noindent$^*$ Helmholtz \cite{helmholtz1863} introduced the beating-based
dissonance curve that Plomp \& Levelt \cite{plomplevelt1965} later parametrised
from listener data; the two models produce equivalent rankings for our
purposes.

\medskip\noindent
Allowing separate numerator and denominator weights for the primes 2, 3, 5
yields perfect rank agreement ($\rho = 1.000$) on this dataset, but with
six free parameters fitted to 13 data points this likely reflects
interpolation rather than structure. The zero-parameter formula
$f(p/q) = p + \Os(q)$ achieves almost the same correlation ($\rho = 0.989$).


\section{Limitations: the Major Third and Major Sixth Tie}

No simple formula in this class distinguishes the major third ($5/4$, human
rank~5) from the major sixth ($5/3$, human rank~7). Both have $p = 5$ and
$\Os(q) = 2$ (since $\Os(4) = \Os(2^2) = 2$ and $\Os(3) = 2$). The
distinction requires knowing that $q = 4 = 2^2$ is simpler than $q = 3$
under octave equivalence, yet the prime-weighting function cannot see this.

The Tenney height $H(p/q) = p \cdot q$ \cite{tenney1988} --- the standard
complexity measure in just intonation theory --- makes the situation worse:
$H(5/4) = 20 > H(5/3) = 15$, so Tenney ranks the major sixth as
\emph{more consonant} than the major third, the opposite of human judgments.
Gradus, $f$, and $\max$ all tie the pair; Tenney inverts them.
This failure of Tenney height is a direct consequence of its symmetric and
multiplicative structure: $H(p/q) = p \cdot q$ is dominated by whichever of
$p$ and $q$ is larger, which for $5/3$ is $q = 3$ (small) rather than for
$5/4$ where $q = 4$ (larger). The asymmetric formula $f$ ties them for the
same structural reason --- $\Os(3) = \Os(4) = 2$ --- but does not
invert the order.

Lahdelma \& Eerola \cite{lahdelma2016} decompose consonance ratings into
sensory and familiarity components; on \emph{sensory consonance} alone,
$5/4$ and $5/3$ are nearly indistinguishable across listeners, suggesting
that the tie in $f(p/q)$ may accurately reflect the purely arithmetic
component of the percept. A subsequent study \cite{lahdelma2021} further
shows that register and inversion affect consonance ratings independently of
roughness and harmonicity, providing direct empirical support for the
asymmetric role of the bass note assumed in Section~7.
The gap observed in Krumhansl's data is plausibly
attributable to learned exposure --- major thirds occur far more frequently
than major sixths in tonal music --- consistent with the finding of McDermott et al.\ \cite{mcdermott2016} that
consonance preferences in a culture with minimal Western music exposure are
substantially reduced. The converse view --- that consonance preferences
reflect universal statistics of harmonic sounds in the environment, including
voiced speech --- is advanced by Bowling, Purves \& Gill \cite{bowling2018},
who find that vocal similarity predicts interval attraction across cultures;
on this account the residual consonance of ET thirds may reflect their
partial alignment with vocal harmonic spectra.
Resolving the tie within an arithmetic formula would therefore require
modelling cultural familiarity, which is outside the scope of the present
approach.


\section{A Two-Stage Perceptual Hypothesis}

We propose a speculative two-stage perceptual interpretation of the formula,
as a qualitative hypothesis about how the auditory system may evaluate an
interval. This is a perceptual hypothesis, not a mathematical conjecture:
it cannot be proved from the formula, but it can be tested experimentally.

\medskip\noindent
\textbf{Stage 1 --- Bass establishes harmonic context (cost $\Os(q)$).}
The brain does not literally factorise integers: the auditory system performs
subharmonic template matching or autocorrelation across candidate periods
\cite{parncutt1988,terhardt1979}. The prime decomposition $\Os(q)$ acts as an
\emph{analytical proxy} for this biological process, counting the
octave-reduction and prime-ascent steps needed to reach the implied
fundamental, and mirroring the complexity of the template search without
claiming that the brain performs explicit prime factorisation.

In this model, the lower note is the $q$-th partial of an implied fundamental
$F$. The brain extrapolates downward to $F$ by ``dividing out'' the prime
factors of~$q$: each factor of 2 (an octave) costs 1 unit; each factor of 3
costs~2; each factor of 5 costs~4. This is exactly $\Os(q)$. Intervals whose
bass note is a power of 2 (e.g.\ the octave, $q=1$; the major second bass,
$q=8=2^3$) establish the fundamental cheaply via octave equivalence. Intervals
with an odd prime in $q$ (the fourth, $q=3$; the minor third bass, $q=5$)
require the brain to extrapolate through a genuinely new prime, incurring
higher cost. The subharmonic matching mechanism is formalised by Parncutt
\cite{parncutt1988}, extending Terhardt's virtual pitch theory to chord roots.

\medskip\noindent
\textbf{Stage 2 --- Upper note reaches into the series (cost $p$).}
Once the fundamental is established, the upper note must be recognised as its
$p$-th partial. Higher partials are quieter (amplitude $\propto 1/p$ for
harmonic sounds with a falling spectral envelope, such as bowed strings and
the singing voice; the exact roll-off is instrument-dependent), more narrowly
tuned, more sensitive to mistuning, and have lower pitch salience
\cite{terhardt1979}. The cost of confirming the upper note as a member of the
harmonic series is therefore modelled as $p$ itself.

\medskip\noindent
\textbf{Hypothesised total perceptual effort:} $f(p/q) = p + \Os(q)$.

\medskip\noindent
This hypothesis is consistent with:
\begin{itemize}
  \item \textbf{Galileo} (1638): pulse coincidences as physical mechanism;
  \item \textbf{Rameau} (1722): the \emph{corps sonore} (harmonic series) as
        the source of musical meaning;
  \item \textbf{Euler} (1739): prime arithmetic as the language of harmonic
        complexity;
  \item \textbf{Terhardt} (1979): virtual pitch and harmonic template matching
        as auditory mechanisms;
  \item \textbf{Huron} \cite{huron2001}: statistical learning of harmonic patterns as
        the basis of consonance judgments.
\end{itemize}

\noindent
We note that the formula $p + \Os(q)$ is not the unique arithmetic expression
consistent with this interpretation: other choices such as $p + \log q$ or
$p + \Omega(q)$ (unweighted prime count) are structurally plausible. The
particular fit of $\Os(q)$ --- Euler's prime-weighted function --- to human
data is an empirical observation rather than a theoretical necessity, and
invites further experimental investigation.


\section{The Consonance Landscape and Equal Temperament}

The three formulae are defined on rational numbers, but musical intervals in
practice are never exact. To study how each formula behaves near a just ratio
we sweep the octave in 1-cent steps, find the simplest rational approximation
$p/q$ (denominator $\leq 64$) within $\pm 15$ cents of each position, and
evaluate $G$, $f$, and $\max$. The result is a \emph{consonance landscape}
--- a spiky function of pitch interval in which low values cluster sharply
around simple fractions and rise steeply on either side.

\subsection*{Well depth and sharpness}

Table~\ref{tab:rivals} shows, for each standard just interval, the nearest
rival rational within $\pm 50$ cents and the corresponding jump in each
formula. The wells around simple fractions are deep for all three functions,
but $f(p/q)$ is substantially steeper: moving 2--5 cents off the fifth (3/2)
takes $f$ from 4 to 104 ($+100$), compared with Gradus rising from 4 to 58
($+54$). The major third is even more extreme: 3 cents off 5/4, $f$ leaps
from 7 to 136, while Gradus rises from 7 to 81. This sharper gradient means
$f(p/q)$ suggests tighter perceptual boundaries around just intonation than
Gradus does --- a potentially testable difference.

\begin{table}[h]
\centering
\caption{Nearest rival rational for each just interval and the resulting jump
in each formula. ``Rival'' is the simplest fraction within $\pm 50$ cents
that is distinct from the just ratio; error is the rival's distance in cents.}
\label{tab:rivals}
\smallskip
\begin{tabular}{llrrrrrr}
\toprule
Interval & Just & Rival & Err (\ct{}) & $\Delta G$ & $\Delta f$ & $\Delta \max$ \\
\midrule
Major second  & $9/8$   & $64/57$  & 2 & $+19$ & $+72$  & $+55$ \\
Minor third   & $6/5$   & $77/64$  & 2 & $+15$ & $+73$  & $+71$ \\
Major third   & $5/4$   & $76/61$  & 3 & $+74$ & $+129$ & $+71$ \\
Fourth        & $4/3$   & $83/62$  & 4 & $+109$& $+108$ & $+79$ \\
Tritone       & $45/32$ & $52/37$  & 1 & $+37$ & $+38$  & $+7$  \\
Fifth         & $3/2$   & $94/63$  & 5 & $+54$ & $+100$ & $+91$ \\
Minor sixth   & $8/5$   & $101/63$ & 2 & $+103$& $+99$  & $+93$ \\
Major sixth   & $5/3$   & $103/62$ & 3 & $+127$& $+127$ & $+98$ \\
Minor seventh & $9/5$   & $16/9$   & 1 & $0$   & $+7$   & $+7$  \\
Major seventh & $15/8$  & $118/63$ & 1 & $+60$ & $+110$ & $+103$ \\
\bottomrule
\end{tabular}
\end{table}

\subsection*{Equal temperament}

Twelve-tone equal temperament (12-TET) divides the octave into 12 equal
semitones of 100 cents each; no interval except the unison and octave is
exactly just. The pattern is clear: the fourth and fifth have ET errors of
only 2 cents and low formula values; the thirds and sixths have errors of
14--16 cents and moderate values; the minor second and tritone have both
large errors and the highest complexity values. Each ET semitone relies on
the ear pulling a
mistuned pitch into the low-complexity well of its just target --- the same
well structure examined in Section~5 (\emph{Comparison of the Three Formulae}).

\subsection*{Alternative equal temperaments}

The tolerance model makes a natural prediction about which equal divisions of
the octave (EDOs) are acoustically viable for which intervals. An $n$-EDO
divides the octave into $n$ equal steps of $1200/n$ cents; an interval is
\emph{well-approximated} when the nearest step falls within
$\Delta_{\mathrm{tol}}$ of the just target. The fifth (tolerance $\sim 10$\ct{}
at A3, $\tau = 4$ Hz) is well-approximated by any EDO whose fifth is within
10 cents of 702 cents: this includes 12-TET (2\ct{} error), 19-TET (7\ct{}), 31-TET
(5\ct{}), and 53-TET (0.07\ct{}), but excludes 5-TET (720\ct{}, error 18\ct{}) and
7-TET (686\ct{}, error 16\ct{}). The major third (tolerance $\sim 6$\ct{} at A3) is
well-approximated only by 19-TET (1.5\ct{} error), 31-TET (0.8\ct{}), and 53-TET
(1.4\ct{}); 12-TET (14\ct{}) fails this test at most registers. This arithmetically
explains the longstanding preference of renaissance-era theorists and
contemporary microtonal composers for 19-TET and 31-TET: these systems
preserve 5-limit harmony within the acoustic tolerance budget, whereas 12-TET
can only do so for the fifth and fourth. The framework thus connects the
arithmetic complexity measure $f(p/q)$ to a principled criterion for
evaluating tuning systems.

\subsection*{The minor seventh and competing tunings}

The minor seventh provides a striking special case. The Ptolemaic minor
seventh $9/5$ (996 cents) and the Pythagorean minor seventh $16/9$ (995
cents) are only 1 cent apart, yet have distinct arithmetic characters.
Gradus gives both the same value ($G = 9$), and is therefore unable to
distinguish these historically competing tunings. The formula $f$ does
distinguish them: $f(9/5) = 13$ vs $f(16/9) = 20$, favouring the Ptolemaic
tuning. The Pythagorean version $16/9 = (4/3)^2$ arises as two stacked fourths and
is the minor seventh generated by the Pythagorean system, in which all
intervals derive from the perfect fifth. Its higher $f$ value reflects its
larger numerator: the upper note sits at the 16th partial, far higher in the
harmonic series than the 9th partial of the Ptolemaic version. Gradus, being
symmetric in $p$ and $q$, cannot see this distinction ($\Os(16) + \Os(9) =
\Os(9) + \Os(16)$); the asymmetry of $f$ is what exposes it.


\section{A Partial-Beating Model of Perceptual Tolerance}

The preceding section treats tolerance as a simple frequency deviation in
cents. We now propose a qualitative model connecting partial-beating rates to
the arithmetic structure of the interval. The components of this model ---
the connection between beating and roughness (\cite{helmholtz1863};
\cite{plomplevelt1965}), the role of virtual pitch \cite{terhardt1979}, and
the perceptual salience of partials \cite{parncutt1988} --- are individually
well-established. The derivation of a $1/p$ tolerance scaling from the formula
$f(p/q)$ is, to our knowledge, new, and should be understood as a
qualitative conjecture pending experimental test.

\subsection*{The model}

For a just interval $p/q$, the $m$-th coincident partial occurs at frequency
$mp \cdot f_0$ (the $(mp)$-th harmonic of the lower note, with amplitude
$\propto 1/(mp)$). If the actual interval deviates by $\Delta$ cents, the
beat rate at the $m$-th coincidence is:
\[
  \beta_m \approx mp \cdot f_0 \cdot \frac{\ln 2}{1200} \cdot |\Delta|
  \quad \text{Hz}
\]
The total roughness contribution, weighted by partial amplitude $\propto
1/(mp)$, is:
\[
  R \propto \sum_{m=1}^{\infty} \frac{1}{mp} \cdot \beta_m
  = \frac{f_0 \ln 2}{1200} \cdot |\Delta| \cdot \sum_{m=1}^{M} 1
  = \frac{f_0 \ln 2}{1200} \cdot |\Delta| \cdot M
\]
where the $mp$ factors cancel. Summing over all coincidences removes the dependence on $p$ under the
simplifying assumptions of this model. Empirically, however, lower-order
coincidences are more perceptually salient than higher ones
\cite{terhardt1979,parncutt1988}, motivating a first-coincidence
approximation in which tolerance scales as $1/p$.

Setting a threshold $\tau$ Hz on the first-coincidence beat rate gives a
tolerance half-width:
\[
  \boxed{\Delta_{\mathrm{tol}}(p/q,\, f_0) \approx \frac{1730\,\tau}{p \cdot f_0}
  \quad \text{cents}}
\]
The Plomp--Levelt critical bandwidth at frequency $f$ is approximately
$\mathrm{CB}(f) \approx 1.72 f^{0.65}$ Hz \cite{plomplevelt1965}; roughness
is first perceptible at ${\sim}0.05\,\mathrm{CB}$ and reaches its maximum
near $0.25\,\mathrm{CB}$. At the first coincidence of the fifth ($p=3$,
frequency $3 \times 220 = 660$ Hz at A3), $\mathrm{CB}(660) \approx 118$ Hz,
giving roughness onset at ${\sim}6$ Hz. This calibrates the three thresholds
used below: $\tau = 4$ Hz (below roughness onset, barely perceptible),
$\tau = 8$ Hz (perceptible beating, not yet rough), and $\tau = 15$ Hz
(at or above roughness onset for most registers).

Two properties follow. First, \textbf{tolerance is proportional to $1/p$}:
the same numerator that appears in $f(p/q)$ determines how precisely the
interval must be tuned to avoid audible beating. Consonant intervals (small
$p$, small $f$) are not only perceived as smooth but are also forgiving of
mistuning. Second, \textbf{tolerance depends on register}: at higher pitches
the tolerance in cents narrows because the beating partials are at higher
absolute frequencies.

\subsection*{Equal temperament in the model}

\begin{figure}[h]
\centering
\includegraphics[width=\textwidth]{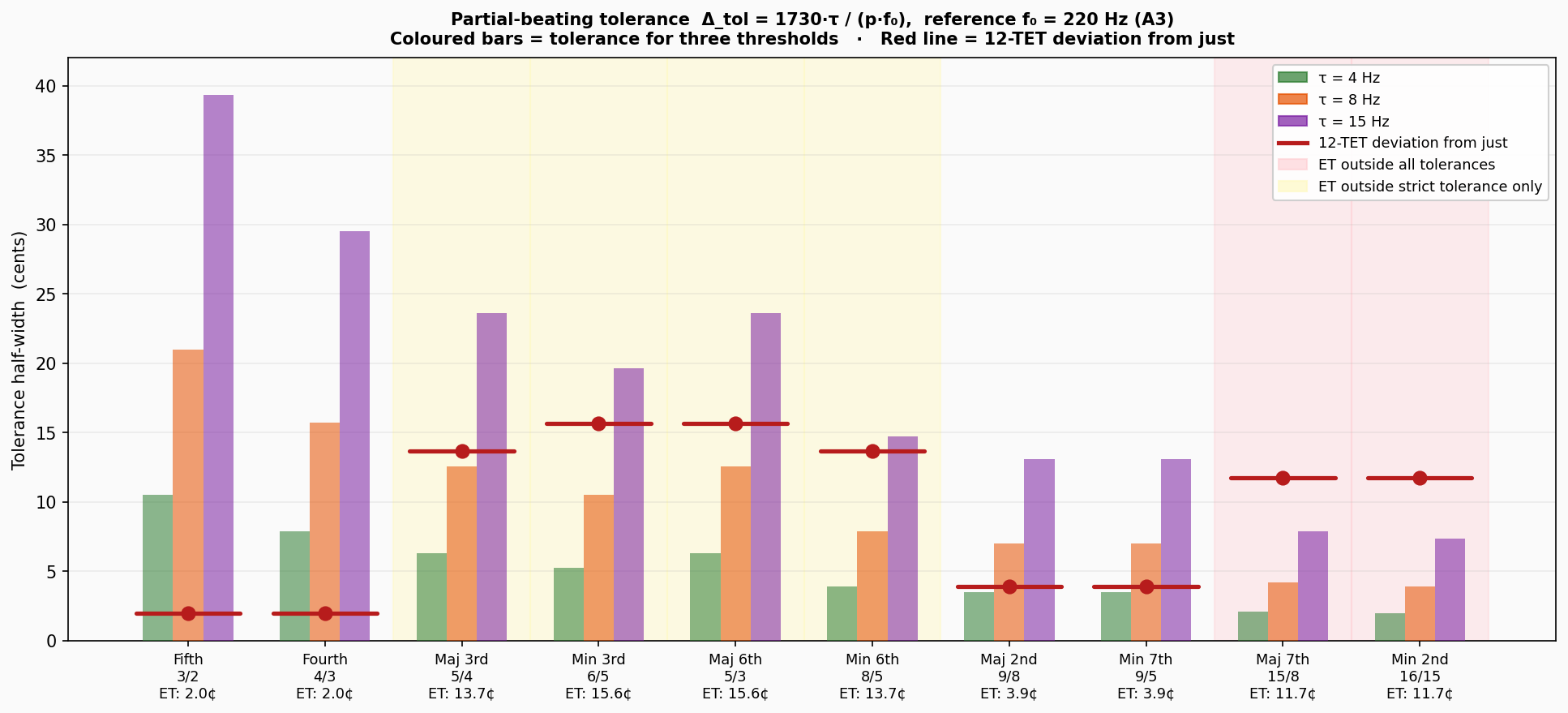}
\caption{Partial-beating tolerance half-width vs 12-TET deviation for each
just interval, at $f_0 = 220$ Hz (A3). Coloured bars show tolerance for
three threshold values; the red mark is the 12-TET deviation. Yellow
shading: ET exceeds the strict (4 Hz) but not the generous (15 Hz)
threshold. Pink shading: ET exceeds all three thresholds.}
\label{fig:tol1}
\end{figure}

Figure~\ref{fig:tol1} shows the tolerance half-width for each standard
interval at a reference register of $f_0 = 220$ Hz (A3), for three threshold
values $\tau \in \{4, 8, 15\}$ Hz. The 12-TET deviation from the just ratio
is shown as a red horizontal mark.

\noindent
The result is unambiguous for the fifth (3/2) and fourth (4/3): their ET
deviations of 2 cents are well within tolerance at all three thresholds and
at all musically relevant registers. The thirds, sixths, and sevenths fall in
a different category: their ET deviations of 14--16 cents substantially
exceed the strict (4 Hz) threshold and sit near or above the generous (15 Hz)
threshold. The major and minor seconds occupy an intermediate position.
The major seventh (15/8) and minor second (16/15) are outside all three
tolerance bands: the large numerator $p \in \{15, 16\}$ means even a modest
deviation produces rapid beating.

\begin{figure}[h]
\centering
\includegraphics[width=0.92\textwidth]{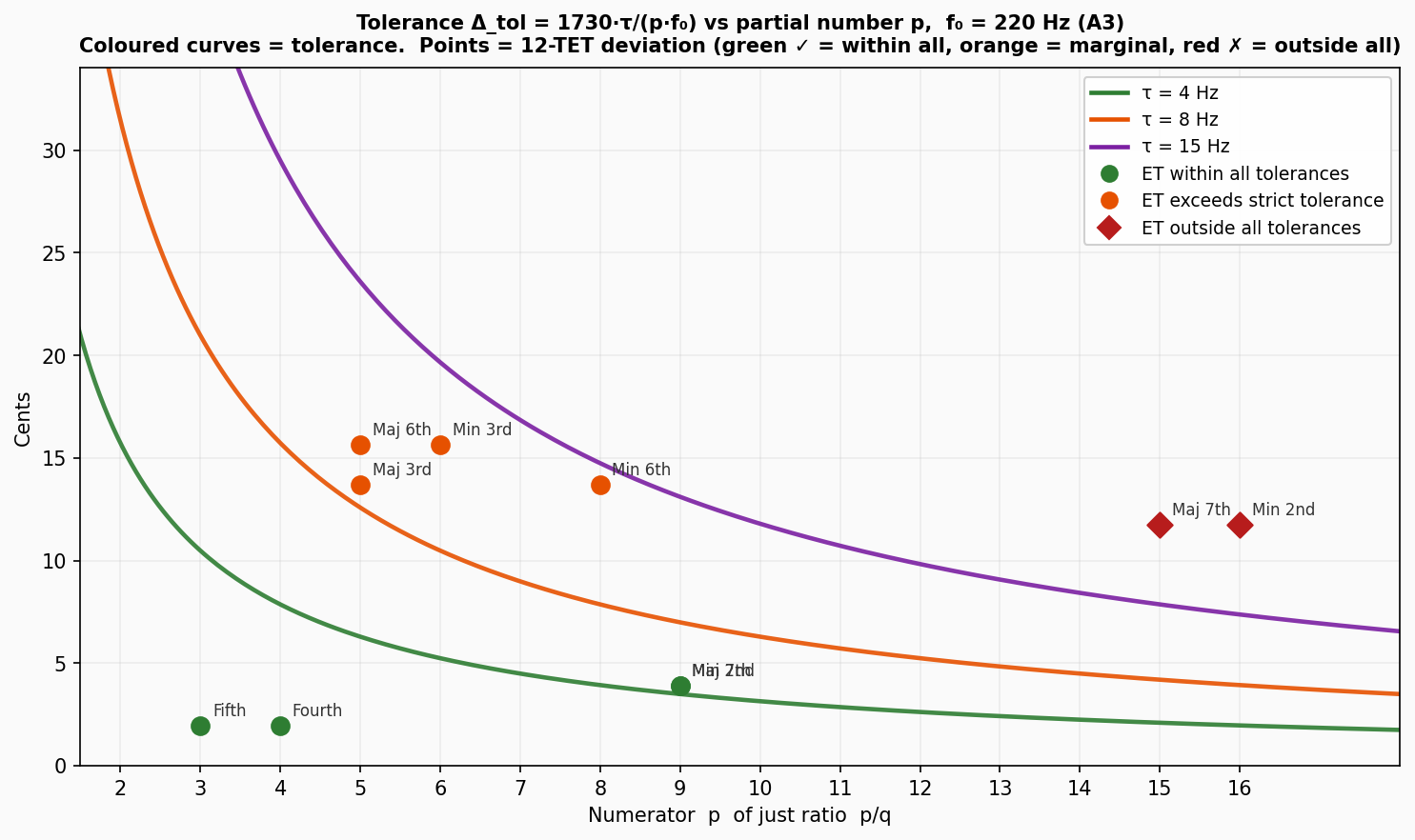}
\caption{Tolerance half-width $\Delta_{\mathrm{tol}} = 1730\tau/(p \cdot f_0)$
as a function of the partial number $p$ at $f_0 = 220$ Hz. Coloured curves
show three threshold values. Interval points show their 12-TET deviations:
green = ET within all tolerances, orange = ET exceeds strict tolerance,
red = ET outside all tolerances.}
\label{fig:tol3}
\end{figure}

Figure~\ref{fig:tol3} plots the tolerance curves directly as functions of
$p$, showing the $1/p$ hyperbolic structure. The interval points are coloured
green (ET within all tolerances), orange (ET marginal), or red (ET outside
all tolerances). The distribution confirms the pattern: the two green
intervals --- the fifth and fourth --- are exactly those Western harmony has
used as structural pillars throughout all periods of common practice, while
the orange and red intervals are precisely those that show the greatest
variation across tuning systems and historical periods.

\subsection*{Register dependence}

\begin{figure}[h]
\centering
\includegraphics[width=\textwidth]{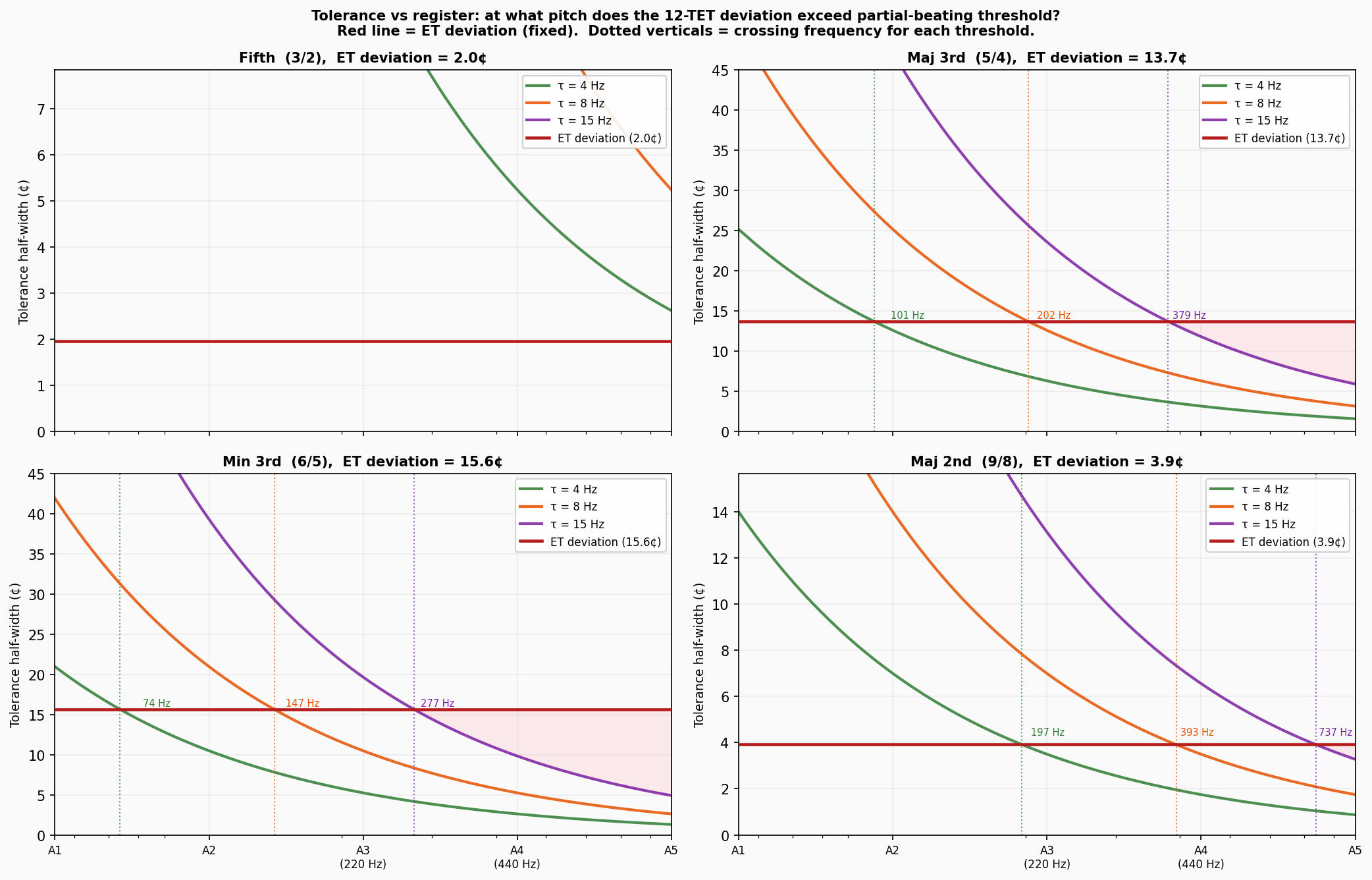}
\caption{Tolerance half-width as a function of register (fundamental
frequency, log scale) for four intervals. The red horizontal line is the
fixed ET deviation; the coloured curves are the three threshold tolerances.
Dotted verticals mark the register at which ET deviation crosses each
threshold. Pink shading indicates the register range where ET exceeds the
generous (15 Hz) threshold.}
\label{fig:tol2}
\end{figure}

Figure~\ref{fig:tol2} shows how tolerance varies with register for four
intervals, with vertical dotted lines marking the fundamental frequency at
which the ET deviation crosses each threshold. For the fifth (top left), the
crossing frequency exceeds 1000 Hz for the moderate threshold --- in practice
the ET fifth is acoustically tolerable throughout the entire playing range of
most instruments. For the major third (top right), the 15 Hz crossing occurs
at around 380 Hz (roughly G4), meaning that in the upper half of the piano
keyboard the ET major third produces beating comfortably above 15 Hz at the
first coincident partial. For the minor third the situation is similar. The
model therefore suggests that ET thirds may sound noticeably rougher in the
upper register than the lower --- a prediction consistent with the common
practice of voicing thirds in the middle-to-lower register in orchestral
writing and avoiding them in high woodwind chords.

\subsection*{The $1/p$ duality}

The model suggests a duality between the two uses of $p$ in this framework.
In the dissonance formula $f(p/q) = p + \Os(q)$, the term $p$ is the
\emph{cost} of the interval: the effort required to identify the upper note
as the $p$-th partial of an implied fundamental. In the tolerance model, $p$
appears in the denominator of $\Delta_{\mathrm{tol}}$: it is the
\emph{precision required} to keep the partial beating within an acceptable
rate. These two roles of $p$ are consistent with the same physical fact ---
that $p$ is the partial number of the upper note in the harmonic series, and
that both the cognitive effort of recognition and the acoustic sensitivity
to mistuning may scale with partial height.

The implication for equal temperament is that Western musical culture has, over
several centuries, adopted a tuning system in which the acoustically robust
intervals (fifth and fourth, small $p$) are preserved almost exactly, while
the acoustically fragile intervals (thirds and sixths, moderate $p$) are
mistuned to the edge of --- and in many registers slightly beyond --- their
acoustic tolerance. Whether this represents a deliberate optimisation,
cultural habituation, or a combination of the two is a question that the
arithmetic model alone cannot resolve.


\section{The Triangle and OEIS Sequences}

The asymmetric formula $f(p/q) = p + \Os(q)$ extends naturally from a
function on pairs to a triangular array of integers indexed by coprime pairs,
and gives rise to two OEIS entries, A397104 and A397106.

\subsection*{The triangle \texorpdfstring{$T(n,k)$}{T(n,k)}}

The formula $f(p/q) = p + \Os(q)$ defines a triangular array of positive
integers. Formally, $T(n,k) = n + \Os(k)$ for $1 \leq k \leq n$ with
$\gcd(n,k) = 1$, together with $T(1,1) = 1$ for the unison. In OEIS
notation this is \emph{the triangle of $n + \Os(k)$ read by rows, where for
each row~$n$ the index~$k$ ranges over integers satisfying $1 \leq k \leq n$
and $\gcd(n,k) = 1$}; only coprime pairs are listed, so rows have no blank
entries and the row length equals $\varphi(n)$ for $n \geq 2$.
The triangle begins:
\[
\begin{array}{rllllll}
n=1: & 1 \\
n=2: & 2 \\
n=3: & 3, & 4 \\
n=4: & 4, & 6 \\
n=5: & 5, & 6, & 7, & 7 \\
n=6: & 6, & 10 \\
n=7: & 7, & 8, & 9, & 9, & 11, & 10 \\
n=8: & 8, & 10, & 12, & 14 \\
n=9:  & 9,  & 10, & 11, & 13, & 15, & 12 \\
n=10: & 10, & 12, & 16, & 14 \\
\end{array}
\]
Here $\varphi$ is Euler's totient function (OEIS A000010).
The flat sequence of terms read left-to-right by rows begins:
\begin{center}
\texttt{1, 2, 3, 4, 4, 6, 5, 6, 7, 7, 6, 10, 7, 8, 9, 9, 11, 10, \ldots}
\end{center}
The connection to Euler's Gradus (A275314) is:
$T(n,k) = n + \mathrm{A275314}(k) - 1$.
The symmetric version of the triangle, $\mathrm{A275314}(n) +
\mathrm{A275314}(k) - 1$, recovers Gradus. This triangle is catalogued
as OEIS A397104.

The rightmost populated diagonal of the triangle, $T(n,\,n-1)$ for
$n \geq 2$, gives the two-stage dissonance of the superparticular
(epimoric) interval $n/(n-1)$ --- the sequence of intervals between
consecutive partials of the harmonic series: octave (2/1), fifth (3/2),
fourth (4/3), major third (5/4), \ldots\ The diagonal sequence begins:
\begin{center}
\texttt{2, 4, 6, 7, 10, 10, 14, 12, 14, 16, 22, 17, 26, \ldots}
\end{center}
Note that $T(8,7) = 14 > T(9,8) = 12$: the sequence is not monotone,
because $\Os(8) = \Os(2^3) = 3$ is small despite $8$ being large, making
the major second $9/8$ cheaper than the septimal interval $8/7$.
This diagonal sequence is catalogued as OEIS A397106, indexed there by
the lower partial $m = n-1$ rather than the upper partial $n$, giving
the equivalent formula $a(m) = m + \mathrm{A275314}(m)$ for $m \geq 1$.

Several companion sequences arise naturally from the triangle and are noted
here for potential future OEIS entries. The \emph{row sums}
$S(n) = \sum_{\substack{k=1\\\gcd(n,k)=1}}^{n} T(n,k)
       = n\,\varphi(n) + \sum_{\substack{k=1\\\gcd(n,k)=1}}^{n} \Os(k)$
measure the total two-stage dissonance across all coprime intervals with
denominator~$n$; the first sum $n\varphi(n)$ dominates, but the correction
term reflects the prime structure of the coprime residues. The \emph{row
maxima} $\max_{k:\,\gcd(n,k)=1} T(n,k)$ and \emph{row minima}
$\min_{k:\,\gcd(n,k)=1} T(n,k)$ reveal the spread of dissonance values
within each harmonic series level and are sensitive to the prime
factorisation of~$n$: rows where $n$ is prime have $\varphi(n)=n-1$ entries
all of the form $n + \Os(k)$, while rows where $n$ is a prime power have
fewer coprime denominators with characteristically low $\Os(k)$ values.


\subsection*{Concluding remarks}

The main contributions of this note are: (i)~a reinterpretation of Euler's
Gradus Suavitatis as a weighted harmonic coincidence count, under a
discrete model of integer-indexed harmonics, connecting it to Galileo's
pulse-coincidence model; (ii)~the asymmetric formula $f(p/q) = p + \Os(q)$,
which matches the best known rank correlation with human consonance data while
resolving more ties than Gradus or $\max(p,q)$; (iii)~a speculative two-stage
perceptual hypothesis offering a cognitive reading of $f$; (iv)~a
partial-beating tolerance model that links the term $p$ in $f$ to acoustic
sensitivity to mistuning, with implications for equal temperament; and
(v)~the coprime integer triangle $T(n,k)$ and its superparticular diagonal,
catalogued as OEIS A397104 and A397106 respectively.

Several directions suggest themselves for future work. The consonance ratings
of Krumhansl \cite{krumhansl1990} are based on a small sample of trained
Western listeners; testing the formula against larger and more culturally
diverse datasets \cite{mcdermott2016,bowling2018} would establish how far
the arithmetic structure of $f$ generalises beyond that context. The
perceptual hypothesis of Section~7 calls for direct experimental investigation:
if Stage~1 cost (bass complexity) and Stage~2 cost (upper-note partial height)
are genuinely separable, they should be dissociable by manipulating register,
inversion, and timbre independently. The tolerance model of Section~9 makes a
concrete and testable prediction --- that mistuning sensitivity scales as
$1/p$ at the first coincident partial --- which could be evaluated with
psychoacoustic paradigms. Finally, the arithmetic structure of the triangle
$T(n,k)$ invites number-theoretic investigation: properties of the row sums,
maxima, and minima remain largely unexplored.


\subsection*{Acknowledgements}

The author thanks E.\ Keith Lloyd (University of Southampton), who
supervised an undergraduate project on mathematics and the theory of
music in 1983--84 and first introduced him to Euler's Gradus Suavitatis.


\appendix

\section{Code}
\label{app:code}

All tables and figures in this paper were produced using Python; the complete
code is available at \url{https://github.com/davidderoure/gradus}.

The following implementations define $\Os(n)$, Euler's Gradus $\G(p/q)$,
the asymmetric formula $f(p/q) = p + \Os(q)$, $\max(p,q)$, the
OEIS triangle $T(n,k)$, and its superparticular diagonal as catalogued in
OEIS A397106 (Section~10). Each is self-contained and sufficient to
verify any value in the paper or generate OEIS sequences.

\subsection*{PARI/GP}

\begin{verbatim}
\\ Omega*(n) = sum of e*(p-1) over prime-power factors p^e || n
OmegaStar(n) = {
  if(n <= 1, return(0));
  my(f = factor(n));
  sum(i = 1, #f~, f[i,2] * (f[i,1] - 1))
};

\\ Euler's Gradus Suavitatis G(p/q)
Gradus(p, q) = {
  my(g = gcd(p,q)); p /= g; q /= g;
  1 + OmegaStar(p) + OmegaStar(q)
};

\\ Asymmetric formula f(p/q) = p + Omega*(q)
f(p, q) = {
  my(g = gcd(p,q)); p /= g; q /= g;
  p + OmegaStar(q)
};

\\ Galileo / Farey metric max(p,q)
maxpq(p, q) = {
  my(g = gcd(p,q)); p /= g; q /= g;
  max(p, q)
};

\\ OEIS triangle T(n,k): row n, column k (1 <= k <= n, gcd(n,k)=1)
T(n, k) = if(gcd(n,k) == 1, f(n,k), 0);
tabl(nn) = for(n=1,nn, for(k=1,n, if(gcd(n,k)==1, print1(T(n,k),", "))))

\\ A397106: superparticular diagonal a(m) = m + A275314(m), m >= 1
\\ (m is the lower partial of the interval (m+1)/m)
a(m) = m + 1 + OmegaStar(m);
vector(45, m, a(m))
\end{verbatim}

\subsection*{Mathematica}

\begin{verbatim}
(* Omega*(n) = Sum of e*(p-1) over prime-power factors p^e || n *)
OmegaStar[1] = 0;
OmegaStar[n_Integer /; n > 1] :=
  Total[#[[2]] (#[[1]] - 1) & /@ FactorInteger[n]];

(* Euler's Gradus Suavitatis G(p/q) *)
Gradus[p_, q_] := With[{r = p/q // Numerator, s = p/q // Denominator},
  1 + OmegaStar[r] + OmegaStar[s]];

(* Asymmetric formula f(p/q) = p + Omega*(q) *)
f[p_, q_] := With[{r = p/q // Numerator, s = p/q // Denominator},
  r + OmegaStar[s]];

(* Galileo / Farey metric max(p,q) *)
MaxPQ[p_, q_] := With[{r = p/q // Numerator, s = p/q // Denominator},
  Max[r, s]];

(* OEIS triangle T(n,k): first nn rows *)
T[n_, k_] /; GCD[n,k] == 1 := f[n, k];
Flatten @ Table[T[n,k], {n, 1, 9}, {k, 1, n}, GCD[n,k] == 1]

(* A397106: superparticular diagonal a(m) = m + A275314(m), m >= 1 *)
(* (m is the lower partial of the interval (m+1)/m) *)
a[m_] := m + 1 + OmegaStar[m];
Table[a[m], {m, 1, 45}]
\end{verbatim}

\subsection*{Python}

\begin{verbatim}
from math import gcd
from sympy import factorint  # pip install sympy; or replace with trial division

def omega_star(n):
    """Omega*(n) = sum of e*(p-1) over prime-power factors p^e || n."""
    if n <= 1: return 0
    return sum(e * (p - 1) for p, e in factorint(n).items())

def gradus(p, q):
    """Euler's Gradus Suavitatis G(p/q)."""
    g = gcd(p, q); p, q = p // g, q // g
    return 1 + omega_star(p) + omega_star(q)

def f(p, q):
    """Asymmetric formula f(p/q) = p + Omega*(q)."""
    g = gcd(p, q); p, q = p // g, q // g
    return p + omega_star(q)

def maxpq(p, q):
    """Galileo / Farey metric max(p,q)."""
    g = gcd(p, q); p, q = p // g, q // g
    return max(p, q)

# OEIS triangle T(n,k): first nine rows
for n in range(1, 10):
    row = [f(n, k) for k in range(1, n + 1) if gcd(n, k) == 1]
    print(row)

# A397106: superparticular diagonal a(m) = m + A275314(m), m >= 1
# (m is the lower partial of the interval (m+1)/m)
def a(m):
    return m + 1 + omega_star(m)

print([a(m) for m in range(1, 46)])
\end{verbatim}


\end{document}